\documentclass[structabstract]{aa}  
\usepackage{graphicx}
\usepackage{txfonts}
\begin{document}
   \title{Radio interferometric observations of two core-dominated triple radio sources at $z>3$}

   \author{D.~Cseh\inst{1}
           \and
           S.~Frey\inst{2,4}
           \and
           Z.~Paragi\inst{3,4}
	   \and
	   L.I.~Gurvits\inst{3,5}
           \and
	   K.\'E.~Gab\'anyi\inst{2,4}
          }

   \institute{Laboratoire Astrophysique des Interactions Multi-echelles (UMR 7158),
              CEA/DSM-CNRS-Universit\'e Paris Diderot,\\
              CEA Saclay, F-91191 Gif sur Yvette, France\\
              \email{david.cseh@cea.fr}
         \and
              F\"OMI Satellite Geodetic Observatory, P.O. Box 585, H-1592 Budapest, Hungary\\
              \email{frey@sgo.fomi.hu, gabanyik@sgo.fomi.hu}
         \and
              Joint Institute for VLBI in Europe, Postbus 2, 7990 AA Dwingeloo, The Netherlands\\
              \email{zparagi@jive.nl, lgurvits@jive.nl}
         \and
              MTA Research Group for Physical Geodesy and Geodynamics, P.O. Box 91, H-1521 Budapest, Hungary
          \and
           Institute of Space and Astronautical Science, Japan Aerospace Exploration Agency, 3-1-1 Yoshinodai Chuo-ku,\\
           Sagamihara, Kanagawa 252-5210, Japan.}

   \date{Received June 03, 2010; accepted August 05, 2010}

 
  \abstract
   {}
   {We selected two radio quasars (J1036+1326 and J1353+5725) based on their 1.4-GHz radio structure, which is dominated by a bright central core and a pair of weaker and nearly symmetric lobes at $\sim$10$\arcsec$ angular separation. They are optically identified in the Sloan Digital Sky Survey (SDSS) at spectroscopic redshifts $z>3$. We investigate the possibility that their core-dominated triple morphology can be a sign of restarted radio activity in these quasars, involving a significant repositioning of the radio jet axis.}
   {We present the results of high-resolution radio imaging observations of J1036+1326 and J1353+5725, performed with the European Very Long Baseline Interferometry (VLBI) Network (EVN) at 1.6~GHz. These data are supplemented by archive observations from the Very Large Array (VLA).We study the large- and small-scale radio structures and the brightness temperatures, then estimate relativistic beaming parameters.}
   {We show that the central emission region of these two high-redshift, core-dominated triple sources is compact but resolved at $\sim$10 milli-arcsecond resolution. We find that it is not necessary to invoke large misalignment between the VLBI jet and the large-scale radio structure to explain the observed properties of the sources.}  
   {}

   \keywords{radio continuum: galaxies -- 
             galaxies: active -- 
             galaxies: jets -- 
             quasars: general -- 
             quasars: individual (J1036+1326) -- 
             quasars: individual (J1353+5725)
             }

   \maketitle


\section{Introduction}
Active galactic nuclei (AGNs) are believed to be powered by mass accretion onto supermassive (from $\sim$10$^6$ up to $\sim$10$^{10}$~$M_\odot$) black holes. However, only a small fraction of them (less than 10\%) appear luminous at radio frequencies. A typical lifetime of a radio-loud quasar is $\sim$10$^6$~yr, so the radio jet activity can be regarded as intermittent in the whole lifetime of an object. 

We use the term ``restarted activity"  for a re-occurence of \textit{radio} activity over the lifetime of a source. This can involve interruption and re-ignition of the radio activity (Kaiser et al. \cite{kaiser}), which can be episodic or not. This does not necessarily involve a complete cessation of the radio activity (Schoenmakers et al. \cite{scho}).

The most spectacular examples of recurrent activity in radio-loud AGNs demonstrate rare double-double (e.g. Lara et al. \cite{lara}; Schoenmakers et al. \cite{so,scho}; Kaiser et al. \cite{kaiser}) or triple-double (Brocksopp et al. \cite{brock}) radio galaxies. In these objects, the inner and outer pairs of lobes have a common centre and are generally well aligned (within $8\degr$). The only double-double radio quasar candidate to date, 4C02.27, has been identified by Jamrozy et al. (\cite{jamrozy}). A recent review of recurrent activity in AGNs, along with a list of known double-double sources has been published by Saikia \& Jamrozy (\cite{saikia}). The two highest redshift objects known with confirmed episodic activity are 3C294 ($z=1.877$) and J1835+6204 ($z=0.51$) (Saikia \& Jamrozy \cite{saikia}). 

A modest fraction of AGNs exhibit triple structures at $\sim$10$\arcsec$ angular separation with fainter, nearly symmetric extended lobes, and relatively bright compact cores. We call these core-dominated triple sources (CDTs; Marecki et al. \cite{marecki}). The cores of these sources are typically unresolved ($<5\arcsec$) in the NRAO\footnote{The National Radio Astronomy Observatory (NRAO) is a facility of the National Science Foundation operated under cooperative agreement by Associated Universities, Inc.} Very Large Array (VLA) Faint Images of the Radio Sky at Twenty-centimeters (FIRST) survey (White et al. \cite{first}) at 1.4~GHz. 
In a 1.4-GHz VLA imaging survey of a complete flux-density limited sample of powerful core-dominated sources (with core flux density $S_{\rm 5\,GHz} > 1$~Jy), Murphy et al. (\cite{murphy}) found that less than 10\% of the sources showed similar triple structures below $z$$\sim$1.5. No CDTs were found above this redshift among their 87 sources.

The CDT structure has been interpreted as a possible signature of once ceased and then restarted activity. It could be coupled with the repositioning of the central radio jet axis (Marecki et al. \cite{marecki}). A plausible explanation is that new infalling material from, e.g., a galaxy merger triggers the re-occurence of accretion. The process could eventually result in the merger of the central supermassive black holes. The coalescence of the black holes might cause a sudden flip in the black hole spin, consequently in the orientation of the jet axis (e.g. Merritt \& Ekers \cite{merritt}; Gergely \& Biermann \cite{gergely}). The jet reorientation could also be caused by the precession of the spin axis. 
According to the standard unified model of radio-loud AGN (Urry \& Padovani \cite{urry}), emission from relativistic jets pointing close to our line of sight will be Doppler-boosted. A change in jet orientation thus may lead to an observed increase in core dominance. Very Long Baseline Interferometry (VLBI) can provide information on the milli-arcsecond scale (mas) properties of these compact jets. 

When an extended radio source lacks fuelling from the central engine, it may eventually become an ultra-steep-spectrum source (R\"{o}ttgering et al. \cite{rott}; De Breuck et al. \cite{debreuck}). Deep, low-frequency surveys that are sensitive to this diffuse emission have not provided clear signatures of earlier activity cycles (Cohen et al. \cite{cohen}; Mack et al. \cite{mack}; Sirothia et al. \cite{siro}). This suggests that restarted activity is a rather rare phenomenon.

Here we report on radio-interferometric observations of two high-redshift quasars, J1036+1326 ($z=3.10$) and J1353+5725 ($z=3.46$). We observed these sources with the European VLBI Network (EVN) at 1.6~GHz. We compared their high-resolution radio images with those obtained from archive VLA data on the arcsecond scale. In Sect.~\ref{sample}, we describe the selection of the two target sources. In Sects.~\ref{observ} and \ref{param}, we present our observations and describe the radio properties of the two quasars. Finally, we estimate the viewing angles and analyse the source parameters in the context of possibly restarted activity (Sect.~\ref{discuss}). Throughout this paper we assume a cosmological model with $H_0=70$~km~s$^{-1}$~Mpc$^{-1}$, $\Omega_{\rm m}$=0.3, and $\Omega_{\Lambda}$=0.7.

\section{Target selection}
\label{sample}

We searched for those compact, high-redshift ($z>3$) radio sources that are unresolved ($<5\arcsec$) in the FIRST survey catalogue\footnote{\tt http://sundog.stsci.edu} (White et al. \cite{first}), that have an integral flux density $S_{\rm 1.4\,GHz}>20$~mJy, and that are optically identified with quasars in the Sloan Digital Sky Survey\footnote{\tt http://www.sdss.org} (SDSS). Their spectroscopic redshifts were taken from the 4th edition of the SDSS Quasar Catalog (Schneider et al. \cite{sdss}) which consists of objects in the SDSS 5th Data Release (DR5). The ultimate goal was to define a sample of radio-loud, optically identified, high-redshift quasars that are good candidates for VLBI imaging observations. Earlier experience gained from the Deep Extragalactic VLBI-Optical Survey (DEVOS; Mosoni et al. \cite{mosoni}; Frey et al. \cite{frey}) suggests that the majority (nearly 90\%) of the carefully pre-selected compact FIRST-SDSS objects show compact mas-scale radio emission and thus are suitable targets for VLBI imaging observations at centimetre wavelengths, regardless of spectral index. 

We found two peculiar objects (J1036+1326 and J1353+5725) in the sample of $\sim$100 compact high-redshift radio quasars selected using the method above. Although they are catalogued as close pairs of separate, unresolved sources in FIRST, the analysis of archival 1.4-GHz VLA A-array images clearly revealed extended triple structures (see Sect.~\ref{observ}). Both of the bright quasars are associated with weaker two-sided radio lobes; i.e., they satisfy the criteria for CDT sources. We did not apply any selection based on the radio spectrum.

\section{Observations and data analysis}
\label{observ}

\subsection{VLA observations}
\begin{figure}
\resizebox{8.5cm}{!}{\rotatebox{270}{
\includegraphics{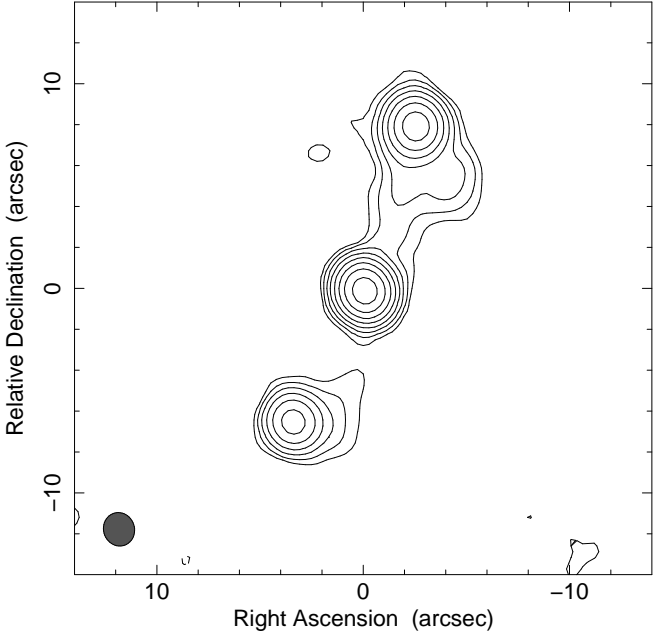}}}
\caption{The naturally-weighted 1.4-GHz VLA A-array image of J1036+1326. The first contours are drawn at $\pm0.25$~mJy/beam. The positive contour levels increase by a factor of 2. The peak brightness is 48 mJy/beam. The Gaussian restoring beam displayed in the lower-left corner is $1\farcs6 \times 1\farcs5 $ (full width at half maximum, FWHM) at a major axis position angle PA=$17\degr$.}
\label{vla1036} 
\end{figure}

\begin{figure}
\resizebox{8.5cm}{!}{\rotatebox{270}{
\includegraphics{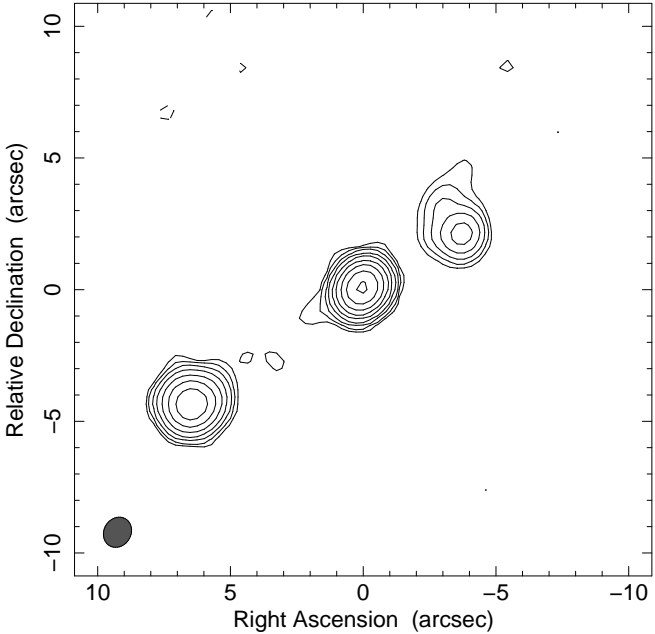}}}
\caption{The naturally-weighted 1.4-GHz VLA A-array image of J1353+5725. The first contours are drawn at $\pm0.6$~mJy/beam. The positive contour levels increase by a factor of 2. The peak brightness is 172 mJy/beam. The Gaussian restoring beam is $1\farcs2 \times 1\farcs0$ at PA=$-29\degr$.}
\label{vla1353} 
\end{figure}

The quasar J1036+1326 was observed with the VLA on 1991 July 8 (project code: AT126). The source J1353+5725 was observed on 1991 June 29 (project code: AR250). For both experiments, the array was in its most extended A configuration, which provided angular resolution of $1\arcsec-1\farcs6$. The observing frequency was 1.4~GHz. The total on-source integration times were 16~min and 2.5~min for J1036+1326 and J1353+5725, respectively. The data sets were obtained from the NRAO Data Archive\footnote{\tt http://archive.nrao.edu}. For spectral information, we also obtained and analysed archive 8-GHz VLA observations of J1036+1326 (project AR415; 1999 August 9) and 5-GHz VLA observations of J1353+5725 (project AW748; 2008 October 15). The on-source integration time in these two last experiments were 1~min and 9.5~min, respectively. The data calibration was performed using the NRAO Astronomical Image Processing System (AIPS; e.g. Greisen \cite{greis}) in a standard way. The 1.4-GHz VLA images of the two sources are shown in Figs.~\ref{vla1036} and \ref{vla1353}.

\subsection{EVN observations}

To check whether the VLA cores of J1036+1326 and J1353+5725 are resolved (for example, show a small separation double structure, thus indicating
a restarting AGN activity) or unresolved (thus showing evidence of a compact jet) on 10 milli-arcsecond scales, we carried out EVN observations at
1.6 GHz. The 6-hour observations were accommodated in the 2009 September 10-11 e-VLBI run. The data were transferred from the telescopes in real-time
to the EVN Data Processor at the Joint Institute for VLBI in Europe (JIVE) in Dwingeloo, the Netherlands (Szomoru \cite{szomoru}). The participating
VLBI stations were  Effelsberg (Germany), Jodrell Bank Lovell Telescope, Cambridge (United Kingdom), Medicina (Italy), Toru\'n (Poland), Onsala (Sweden),
Sheshan (P.R. China), and the phased array of the Westerbork Synthesis Radio Telescope (WSRT; The Netherlands). The total aggregate
bitrate per station was 512~Mbps. There were eight 8~MHz intermediate frequency channels (IFs) in both left and right circular polarisations.

The sources were observed in phase-reference mode. This allowed us to increase the coherent integration time spent on the target source and thus to improve the sensitivity of the observations. Phase-referencing involves regularly interleaving observations between the target source and a nearby, bright, and compact reference source (e.g. Beasley \& Conway \cite{beas}). The delay, delay rate, and phase solutions derived for the phase-reference calibrators (J1025+1253 and J1408+5613 in our experiment) were interpolated and applied to the respective target within the target-reference cycle time of 7~min. The target source was observed for 5-min intervals in each cycle. The total accumulated observing time on each target source was 1.6~hours. 

\begin{figure}
\resizebox{8.5cm}{!}{\rotatebox{270}{
\includegraphics{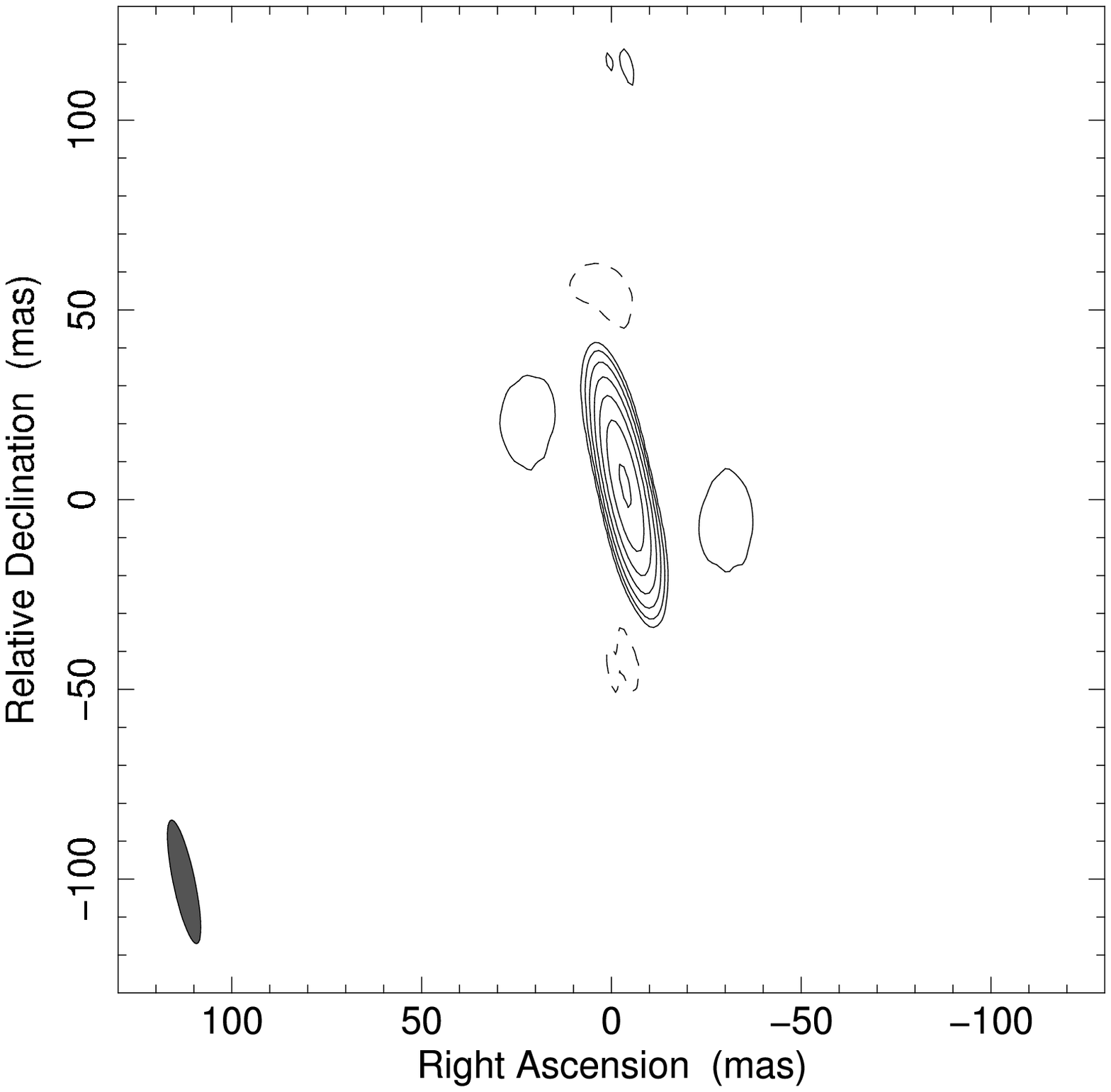}}}
\caption{The 1.6-GHz VLBI image of the core of J1036+1326. The first contours are drawn at $\pm0.71$~mJy/beam. The positive contour levels increase by a factor of 2. The peak brightness is 50 mJy/beam. The Gaussian restoring beam is 33~mas $\times$ 6~mas at PA=$12\degr$.}
\label{evn1036} 
\end{figure}

\begin{figure}
\resizebox{8.5cm}{!}{\rotatebox{270}{
\includegraphics{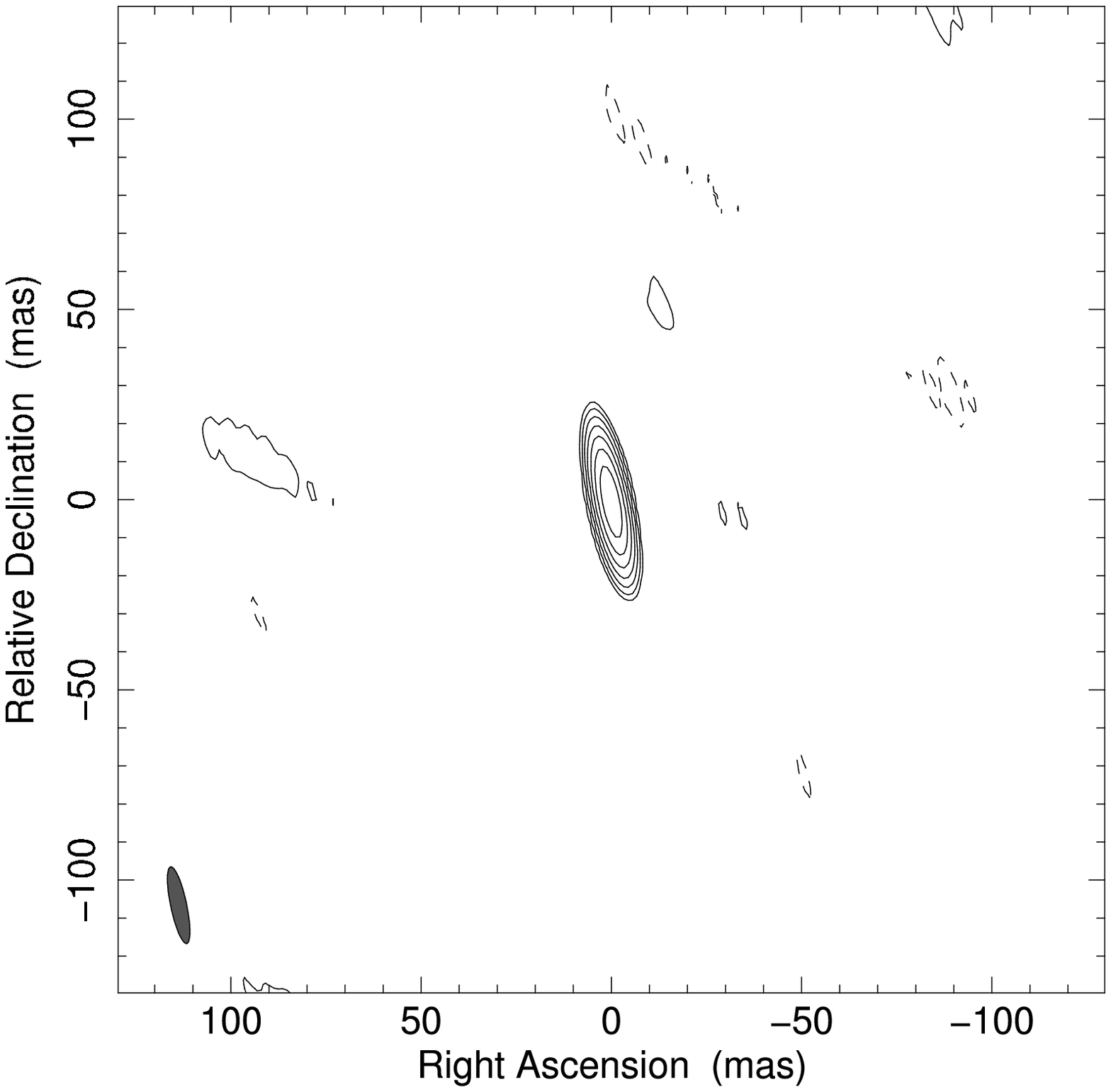}}}
\caption{The 1.6-GHz VLBI image of the core of J1353+5725. The first contours are drawn at $\pm1.7$~mJy/beam. The positive contour levels increase by a factor of 2. The peak brightness is 197 mJy/beam. The Gaussian restoring beam is 21~mas $\times$ 4~mas at PA=$12\degr$.}
\label{evn1353} 
\end{figure}

The AIPS was used for the VLBI data calibration following standard procedures (e.g. Diamond \cite{dia}). The visibility amplitudes were calibrated using system temperatures and antenna gains measured at the antennas. Fringe-fitting was performed for the calibrator sources using 3-min solution intervals. The calibrated visibility data were exported to the Caltech Difmap program (Shepherd et al. \cite{shep}) where the VLBI images (Fig.~\ref{evn1036}-\ref{evn1353}) were made after several cycles of CLEAN and phase-only (later amplitude and phase) self-calibration iterations. 

\section{Results}
\label{param}

Observing in phase-reference mode allowed us to determine accurate coordinates of the target sources. For J1036+1326, we obtained right ascension $10^{\rm h} 36^{\rm m} 26\fs88565$, declination $+13\degr 26\arcmin 51\farcs7598$ with an error of 1 mas. For J1353+5725, we obtained $13^{\rm h} 53^{\rm m} 26\fs03368$, $+57\degr 25\arcmin 52\farcs9097$ with an error of 2 mas. 

The angular extent of the large-scale 1.4-GHz radio structure of J1036+1326 (Fig.~\ref{vla1036}) is $15\farcs8$, which corresponds to a projected linear size of $\sim$120~kpc at $z=3.10$. For the quasar J1353+5725 (Fig.~\ref{vla1353}), the angular size is $12\farcs23$, and the linear size is $\sim$90~kpc at $z=3.46$. 

To characterise the source brightness distributions on the VLA scale with simple models, we used Difmap to fit circular Gaussian model components to the self-calibrated interferometric visibility data. The flux density of the best-fit model components are listed in Table~\ref{VLA}, along with two-point spectral indices derived from the flux densities measured at different frequencies. The spectral index ($\alpha$) was calculated by assuming a power-law dependence of the flux density ($S$) and the observing frequency ($\nu$): $S\sim \nu^{\alpha}$. The flux density errors were estimated following the approach by Fomalont (\cite{fomalont}) and correspond to 1-$\sigma$. The 1-$\sigma$ error of the spectral indices is less than 0.05. 

\begin{table}
\caption{Observed and derived parameters of the two quasars from the VLA measurements}
\label{VLA}
\begin{center}
\begin{tabular}{ccccc}
\hline
Name       & Component   & $S_{\rm 1.4\,GHz}$  & $S_{\rm 5/8\,GHz}$  & $\alpha$ \\
           &             &        [mJy]        & [mJy]          &          \\
\hline
J1036+1326 & Core        &  50.4$\pm$1.1         & 62.6$\pm$1.3            & +0.1    \\
             & North     &       27.8$\pm$0.6         &  2.9$\pm$0.2            & $-$1.2  \\
           & South       &       13.0$\pm$0.4         &  1.7$\pm$0.2            & $-$1.1  \\
\hline
J1353+5725 & Core         &177.9$\pm$3.6         &144.9$\pm$2.9           & $-$0.2  \\
           & North       &        18.6$\pm$0.7         &  3.9$\pm$0.5            & $-$1.3  \\
           & South       &        94.0$\pm$2.0         & 20.9$\pm$0.7            & $-$1.3  \\
\hline
\end{tabular}
\end{center}
\small{
$S_{\rm 1.4\,GHz}$ is the flux density at 1.4~GHz (in the L band), while $S_{\rm 5/8\,GHz}$ denotes the 8~GHz (X-band) flux density in the case of J1036+1326, and the 5~GHz (C-band) flux density in the case of J1353+5725.}
\end{table}

In Table~\ref{EVN}, we list the parameters of the circular Gaussian models (flux density, size) fitted to the 1.6-GHz EVN visibility data in Difmap. We also derived the brightness temperatures according to the following formula (e.g. Condon et al. \cite{condon}):  
\begin{equation}
T_{\rm b}=1.22 \times 10^{12} (1+z) \frac{S}{\vartheta^{2} \nu^{2}}\,\,\, [\mathrm{K}]
\end{equation}
where $S$ is the flux density (Jy), $\nu$ is the observed frequency (GHz) and $\vartheta$ the FWHM size of the Gaussian (mas). 

\begin{table}
\caption{Observed and derived parameters of the two quasar cores from the EVN measurements}
\label{EVN}
\begin{center}
\begin{tabular}{ccccc}
\hline
Name       &Redshift* &$S_{\rm 1.6\,GHz}$ [mJy]** & $\vartheta$ [mas] & $T_{\rm b}$ [K]\\
\hline
J1036+1326 & 3.10&  58.1$\pm$2.9              & 4.7               & $5 \times 10^{9}$   \\
J1353+5725 & 3.46 &231.1$\pm$11.6              & 2.2               & $1 \times 10^{11}$  \\
\hline
\end{tabular}
\end{center}
\small{*4th edition of the SDSS Quasar Catalog (Schneider et al. \cite{sdss})\\
***errors correspond to 1-$\sigma$}
\end{table}

\section{Discussion}
\label{discuss}

\subsection{Large-scale morphologies and viewing angles}

In order to investigate whether the large-scale morphology is dominated by beamed or unbeamed emission, we estimated the arm-length ratio. The latter corresponds to the light travel time for the radio emission morphological components on the approaching and receding sides. We assumed that the observed arm-length ratios are the result of relativistic motions over the lifetime of the sources rather than the difference in the environment on the two sides. The 1.4-GHz VLA images  (Figs.~\ref{vla1036} and \ref{vla1353}) allowed us to determine the arm-length ratios ($R$) of the north and south components. For the sources J1036+1326 and J1353+5725, we get $R_1 = 1.15$ and $R_2 = 1.84$, respectively.

To place approximate limits on the expansion speed and the viewing angle we used
\begin{equation}
\label{beta}
\beta \cos\phi = (R -1)/(R+1), 
\end{equation}
where $\beta=\varv/c$ is the velocity of the plasma blob ($\varv$) expressed in the units of the speed of light ($c$), and $\phi$ is the angle between the approaching 
jet direction and the line of sight eg. (Taylor \& Vermeulen \cite{taylor}). 

We can thus derive a lower limit to $\beta$ (because $\cos\phi\leq1$), and an upper limit to $\phi$ (because $\beta<1$). For J1036+1326, the constraints ($\phi_1<86\degr$ and $\beta_1\geq0.07$) are not very strict, but for J1353+5725 ($\phi_2<73\degr$ $\beta_2\geq0.30$) they are more meaningful. In particular, from the arm-length ratio alone, we can infer that there is mildly relativistic motion in the case of J1353+5725. This method is valid regardless of the specifics of the jet model. It remains valid for the spots or discrete blobs in the head of the jet created by shocks (Schoenmakers et al. \cite{scho}) and for pairs of unequally-sized edge-brightened lobes of an FR-II morphology (Fanaroff \& Riley \cite{fr}).

\subsection{Small-scale morphologies and viewing angles}
As a starting hypothesis, we assume that the inner radio jets that remain compact with VLBI (Figs.~\ref{evn1036} and \ref{evn1353}) are well aligned with the $\sim$100-kpc scale radio structure we see in the VLA images (Figs.~\ref{vla1036} and \ref{vla1353}). Using the range of the allowed viewing angles ($\phi_{1}<86\degr$ and $\phi_{2}<73\degr$) estimated above, we can calculate the Doppler boosting factors ($\delta$) and compare the expected brightness temperatures with our measured values (Table~\ref{EVN}, Column~5), in order to determine whether we need to introduce large misalignments to describe the observed morphology. 

The observed and intrinsic brightness temperatures are related as
\begin{equation}
\delta = T_{\rm b}/T_{\rm b,int},
\end{equation}
where $\delta$ is the Doppler factor.
Here we assume a typical intrinsic brightness temperature of $T_{\rm b,int} \simeq 5 \times 10^{10}$~K, which is valid if energy equipartition holds between the magnetic field and the particles in the radio source (Readhead \cite{red}). We obtain Doppler factors $\delta_1=0.1$ (which in fact means de-amplification) and $\delta_2=2$ for J1036+1326 and J1353+5725, respectively. 

The Doppler boosting factor can be expressed as
\begin{equation}
\label{delta}
\delta  = \frac{1}{\Gamma(1-\beta \cos \phi)},
\end{equation}
where the Lorentz factor is $\Gamma=(1-\beta^2)^{-1/2}$.

For J1036+1326, $\delta_1=0.1$, and even the possible maximum value of $\phi_1=86\degr$ can be made consistent if the bulk speed in the inner jet is $\beta_{\rm inn}=0.9957$. It would correspond to a Lorentz factor $\Gamma \simeq 11$, a value often measured in other radio-loud AGNs (e.g. Lister et al. \cite{lister}). However, models of the inner jet with much higher Lorentz factors, hence smaller viewing angles, should be considered because of the quasar identification of the source. If we allow for a very high $\Gamma\simeq30$, the misalignment would be $\sim$40$\degr$. For comparison, the largest measured misalignment between the inner and outer lobes in 3C293, a double-double radio galaxy, is $\sim$35$\degr$, although this is higher than the typical values (Saikia \& Jamrozy \cite{saikia}, and references therein). In our case, the pc-scale structure detected is insufficient for making definitive conclusions about alignment of pc- and kpc-scale structures.

In the case of J1353+5725, the measured Doppler factor ($\delta_2=2$) implies a maximum of $30\degr$ for the viewing angle because $\sin\phi\leq 1/\delta$ (e.g. Urry \& Padovani \cite{urry}). Thus we cannot assume $\phi_2=72\degr$ for the inner jet, i.e. we cannot use the same approach as in case of the first source. We can construct a set of reasonable physical and geometric parameters which {\it does not require jet repositioning} and is fully consistent with our observations. On the other hand, the largest possible misalignment could reach about 70\degr. 

In addition, we can set a boundary for the outer $\beta$ parameter by considering the outer structure as an FR-II, since the hot spot advance speeds for high-luminosity FR-II radio galaxies is $\beta\la0.3$ (e.g. Liu et al. \cite{liu}; O'Dea et al. \cite{odea}). Even if we use this boundary, we can explain the source without any misalignment; e.g., if the approaching inner and outer jets are both inclined by $\phi=20\degr$ to the line of sight, then a bulk jet speed $\beta_{\rm inn}=0.9903$ would be consistent with the measured Doppler factor in the inner region (Eq.~\ref{delta}). The corresponding Lorentz factor is $\Gamma \simeq 7$. On the other hand, $\beta_{\rm out}=0.31$ and $\phi=20\degr$ would be consistent with the observed arm-length ratio (Eq.~\ref{beta}) of the outer structures of J1353+5725 where the jet flow most likely becomes mildly relativistic. Therefore, the observed morphologies do not require invoking the concept of jet repositioning, although our data cannot rule out the possibility of misaligned inner and outer structures.
 
\subsection{The radio spectra}

We obtained a two-point radio spectral index of each component in our VLA images (Figs.~\ref{vla1036} and \ref{vla1353}) and listed in Column~5 of Table~\ref{VLA}. In general, the steep spectra of the outer components ($\alpha\simeq-1.2$) may indicate radiative losses. The steep-spectrum sources are good candidates for finding fossils from earlier episodic activity (Sirothia et al. \cite{siro}). However, it is not convincing that we see relic lobes in the present case, because, at least for J1036+1326, there is a clear indication of a jet-like feature connecting the core and the northern lobe in the VLA image (Fig.~\ref{vla1036}). In addition, the core dominance of these sources is somewhat expected at high redshift because the relatively high emitted (rest-frame) frequency is $(1+z)$ times the observing frequency. Considering the steep spectra of the radio lobes (Table~\ref{VLA}), if one places these sources at e.g. $z\simeq0.5$, then the (observed) 1.4-GHz flux density of the brighter lobes in these objects would far exceed that of the flat-spectrum core. 

\section{Conclusions}
\label{concl}

We used radio interferometric imaging observations to study the structure of two high-redshift quasars, J1036+1326 ($z=3.10$) and J1353+5725 ($z=3.46$). Archive VLA data show that the sources are extended to $\sim$100~kpc projected linear sizes. They are dominated by a bright central core and a pair of weaker and nearly symmetric lobes. On the other hand, EVN observations at angular resolution that is almost three orders of magnitude higher confirmed that the central jets are compact, at least on $\sim$10~pc linear scales. CDT sources are sometimes interpreted as examples of once ceased and then restarted radio-loud AGNs. It has been claimed that the reignition of their central engine could be coupled with a change in the spin axis of the central accreting, supermassive black hole, hence with the repositioning of the radio jet direction. 

We described a general method for testing the misalignments between the small- and large-scale structures of triple radio sources, using VLA and VLBI imaging observations. We applied this method for two CDT quasars at high redshift ($z>3$), J1036+1326 and J1353+5725. There is no evidence that their morphology would require any misalignment between their characteristic inner and outer radio structures, although misalignments cannot be ruled out by our data.

Considering the radio spectra of the sources we concluded that the core-dominance of these sources can be explained by their high redshift. The relatively high emitted frequency is $(1+z)$ times the observing frequency. Taking the steep spectra of the radio lobes into account, if placed to $z\simeq0.5$, the 1.4-GHz flux density of the brighter lobes in these objects would far exceed that of the flat-spectrum core.

\section*{Acknowledgments}
The EVN is a joint facility of European, Chinese, South African, and other 
radio-astronomy institutes funded by their national research councils. This activity is supported by the European Community Framework Programme
7, Advanced Radio Astronomy in Europe, grant agreement no. 227290. 
The e-VLBI developments in Europe were supported by the EC DG-INFSO funded 
Communication Network Developments project ``EXPReS'', Contract No. 02662.
The research leading to these results has received funding from the
European Community's Seventh Framework Programme (FP7/2007-2013) under
grant agreement number ITN 215212 ``Black Hole Universe'', and the
Hungarian Scientific Research Fund (OTKA, grant no. K72515).

\end{document}